\documentclass[conference]{IEEEtran}
\IEEEoverridecommandlockouts
\IEEEoverridecommandlockouts

\usepackage{cite}
\usepackage{pifont}
\usepackage[table]{xcolor}
\usepackage{multirow}
\usepackage{colortbl}
\usepackage{array}
\usepackage{ragged2e}
\usepackage{amsmath,amssymb,amsfonts}
\usepackage{booktabs}
\usepackage{caption}
\usepackage{graphicx}
\usepackage{textcomp}
\usepackage{tikz}
\usepackage{algorithm}
\usepackage{algpseudocode}
\usepackage{tabularx}
\usepackage{listings}
\usepackage{mdframed}

\usetikzlibrary{arrows.meta,positioning,calc,fit,backgrounds,shapes.geometric,matrix}

\definecolor{headergray}{RGB}{230,230,230}
\definecolor{hyperblue}{RGB}{220,235,255}
\definecolor{successgreen}{RGB}{0,120,0}
\definecolor{failred}{RGB}{180,0,0}
\definecolor{codegray}{RGB}{240,240,240}
\definecolor{LogBlue}{RGB}{25,70,160}
\definecolor{LogGreen}{RGB}{0,120,60}
\definecolor{LogRed}{RGB}{180,30,30}
\definecolor{LogGray}{RGB}{90,90,90}
\definecolor{LogPurple}{RGB}{120,60,160}
\definecolor{LogBG}{RGB}{248,248,248}

\newcommand{\cmark}{\ding{51}}
\newcommand{\xmark}{\ding{55}}

\lstdefinestyle{svsuccess}{
  language=Verilog,
  backgroundcolor=\color{codegray},
  basicstyle=\ttfamily\footnotesize,
  keywordstyle=\color{successgreen}\bfseries,
  commentstyle=\color{successgreen},
  numbers=none,
  numberstyle=\tiny,
  frame=single,
  breaklines=true
}

\lstdefinestyle{svfailure}{
  language=Verilog,
  backgroundcolor=\color{codegray},
  basicstyle=\ttfamily\footnotesize,
  keywordstyle=\color{failred}\bfseries,
  commentstyle=\color{failred},
  numbers=none,
  numberstyle=\tiny,
  frame=single,
  breaklines=true
}

\lstdefinestyle{hhlog}{
  basicstyle=\ttfamily\scriptsize,
  columns=fullflexible,
  keepspaces=true,
  breaklines=true,
  breakatwhitespace=false,
  frame=single,
  framerule=0.3pt,
  rulecolor=\color{black},
  showstringspaces=false,
  xleftmargin=0.6em,
  xrightmargin=0.6em,
  aboveskip=4pt,
  belowskip=2pt,
  backgroundcolor=\color{LogBG},
  linewidth=\columnwidth,
  keywordstyle=\bfseries\color{LogBlue},
  commentstyle=\color{LogGray},
  alsoletter={=,-,.,_,>,<,/,\\,?,:,*,+},
  morekeywords={PHASE,P1,P2,Correctness,SA,DC,iter,T,pick,score,compile,sim,warn,ACCEPT,REJECT,best_score,saved,area,power,wns,SELECTED}
}

\lstdefinestyle{compilerlog}{
  basicstyle=\ttfamily\footnotesize,
  columns=fullflexible,
  keepspaces=true,
  breaklines=true,
  frame=single,
  rulecolor=\color{black},
  showstringspaces=false,
  aboveskip=6pt,
  belowskip=6pt
}

\def\BibTeX{{\rm B\kern-.05em{\sc i\kern-.025em b}\kern-.08em
    T\kern-.1667em\lower.7ex\hbox{E}\kern-.125emX}}
    
\begin{document}

\title{HYPERHEURIST: A Simulated Annealing-Based Control Framework for LLM-Driven Code Generation in Optimized Hardware Design\\
}

\author{

\IEEEauthorblockN{Shiva Ahir}
\IEEEauthorblockA{
Department of ECE\\
Stony Brook University\\
Stony Brook, NY\\
shiva.ahir@stonybrook.edu
}

\and

\IEEEauthorblockN{Prajna Bhat}
\IEEEauthorblockA{
Department of ECE\\
Stony Brook University\\
Stony Brook, NY\\
prajna.bhat@stonybrook.edu
}

\and

\IEEEauthorblockN{Alex Doboli}
\IEEEauthorblockA{
Department of ECE\\
Stony Brook University\\
Stony Brook, NY\\
alex.doboli@stonybrook.edu
}



}

\maketitle

\begin{abstract}
Large Language Models (LLMs) have shown promising progress for generating Register Transfer Level (RTL) hardware designs, largely because they can rapidly propose alternative architectural realizations. However, single-shot LLM generation struggles to consistently produce designs that are both functionally correct and power-efficient. This paper proposes HYPERHEURIST, a simulated-annealing–based control framework that treats LLM-generated RTL as intermediate candidates rather than final designs. The suggested system not only focuses on functionality correctness but also on Power-Performance-Area (PPA) optimization. In the first phase, RTL candidates are filtered through compilation, structural checks, and simulation to identify functionally valid designs. PPA optimization is restricted to RTL designs that have already passed compilation and simulation. Evaluated across eight RTL benchmarks, this staged approach yields more stable and repeatable optimization behavior than single-pass LLM-generated RTL.
\end{abstract}

\begin{IEEEkeywords}
Large Language Models, Automated Code Generation, Heuristic Optimization, RTL Synthesis, Functional Correctness, Design Space Exploration
\end{IEEEkeywords}

\section{Introduction}

Optimization in hardware design has traditionally relied on fixed rules, manually tuned heuristics, and scripts from experienced verification engineers. As designs grow larger and constraints become tighter, these approaches increasingly struggle to scale because they require constant retuning and a lot of domain expertise. Recent work shows that LLMs can change this pattern  by instead of hard coding heuristics, LLMs can generate and adapt them based on feedback from synthesis, simulation, and quality metrics  \cite{ref1,ref2,ref3}. Prior work\cite{ref1,ref4,ref5,ref6} shows this idea across multiple stages of the design flow, which includes high-level synthesis, physical design, placement, buffering, and rewriting. LLMs move beyond simple suggestions and begin to guide search, balance exploration and refinement, and react to performance feedback, such as timing, area, and power.

In spite of this progress, existing approaches arguably remain fragmented \cite{ref2,ref5,ref6,ref14}. Most are limited to a single design stage and use custom prompting or agent structures that do not transfer well across problems. Feedback mechanisms also vary widely, and durability across designs, tools, and technology settings is still limited. This paper introduces HYPERHEURIST as a unifying framework to address these gaps. HYPERHEURIST generates SystemVerilog code for digital hardware by treating heuristic design itself as a learnable process and applies it to RTL hardware optimization. It combines parallel heuristic generation, refinement that is critique-driven, and Simulated Annealing~\cite{ref7} with a simpler targeted pipeline for focused optimization. A contextual mechanism selects the strategy that best fits the problem and feedback, moving toward a more general and adaptive model of heuristic-driven EDA.

The paper has the following structure. Section~II presents related work and motivation. Section~III presents methodology, Section IV describes experimental setup, and Section V discusses results.


\begin{figure}[t]
\centering
\resizebox{0.98\columnwidth}{!}{%
\begin{tikzpicture}[
    font=\footnotesize,
    card/.style={
        rounded corners=7pt,
        draw=black!55,
        line width=0.8pt
    },
    cardtitle/.style={
        font=\bfseries\small,
        align=center,
        text=black,
        yshift=1pt
    },
    cardsub/.style={
        font=\scriptsize\itshape,
        text=black,
        align=center
    },
    framework/.style={
        circle,
        draw=black!65,
        line width=0.4pt,
        minimum size=5.5pt,
        inner sep=0pt
    },
    flabel/.style={
        font=\scriptsize,
        text=black,
        align=left
    },
    whylabel/.style={
        font=\scriptsize\itshape,
        text=black!85,
        align=left,
        text width=2.05cm
    },
    explabel/.style={
        font=\scriptsize,
        text=black,
        align=center
    }
]

\fill[orange!18] (0,0) rectangle (3.5,5.4);
\fill[blue!16]   (4.0,0) rectangle (7.5,5.4);
\fill[green!18]  (8.0,0) rectangle (11.5,5.4);

\draw[card] (0,0) rectangle (3.5,5.4);
\draw[card] (4.0,0) rectangle (7.5,5.4);
\draw[card] (8.0,0) rectangle (11.5,5.4);

\fill[orange!30] (0,4.45) rectangle (3.5,5.4);
\fill[blue!28]   (4.0,4.45) rectangle (7.5,5.4);
\fill[green!30]  (8.0,4.45) rectangle (11.5,5.4);

\draw[card] (0,4.45) rectangle (3.5,5.4);
\draw[card] (4.0,4.45) rectangle (7.5,5.4);
\draw[card] (8.0,4.45) rectangle (11.5,5.4);

\node[cardtitle] at (1.75,4.95) {Search-efficient};
\node[cardtitle] at (5.75,4.95) {Balanced};
\node[cardtitle] at (9.75,4.95) {High-QoR};

\node[cardsub] at (1.75,4.18) {lower search cost};
\node[cardsub] at (5.75,4.18) {cost--quality trade-off};
\node[cardsub] at (9.75,4.18) {higher QoR, often specialized};

\node[explabel,text width=2.7cm] at (1.75,3.45)
{Fast exploration with lighter optimization overhead};

\node[explabel,text width=2.7cm] at (5.75,3.45)
{Moderate search cost with stronger solution quality};

\node[explabel,text width=2.7cm] at (9.75,3.45)
{More aggressive refinement for higher optimization quality};

\node[framework, fill=blue!60] (hls) at (0.62,2.45) {};
\node[flabel, anchor=west] at ($(hls)+(0.32,0.12)$) {HLS-BO-LLM};
\node[whylabel, anchor=north west] at ($(hls)+(0.32,-0.02)$)
{LLM-guided BO for efficient search};

\node[framework, fill=cyan!70] (idse) at (0.62,1.20) {};
\node[flabel, anchor=west] at ($(idse)+(0.32,0.12)$) {iDSE};
\node[whylabel, anchor=north west] at ($(idse)+(0.32,-0.02)$)
{pruning reduces search space};

\node[framework, fill=teal!60] (agents) at (4.62,2.45) {};
\node[flabel, anchor=west] at ($(agents)+(0.32,0.12)$) {LLM-DSE Agents};
\node[whylabel, anchor=north west] at ($(agents)+(0.32,-0.02)$)
{multi-agent improves exploration};

\node[framework, fill=orange!75] (buff) at (4.62,1.20) {};
\node[flabel, anchor=west] at ($(buff)+(0.32,0.12)$) {BUFFALO};
\node[whylabel, anchor=north west] at ($(buff)+(0.32,-0.02)$)
{RL improves optimization quality};

\node[framework, fill=red!60] (evo) at (8.62,2.45) {};
\node[flabel, anchor=west] at ($(evo)+(0.32,0.12)$) {Evo-Placement};
\node[whylabel, anchor=north west] at ($(evo)+(0.32,-0.02)$)
{evolutionary search for high QoR};

\node[framework, fill=violet!65] (aspen) at (8.62,1.20) {};
\node[flabel, anchor=west] at ($(aspen)+(0.32,0.12)$) {ASPEN};
\node[whylabel, anchor=north west] at ($(aspen)+(0.32,-0.02)$)
{formal refinement establishes quality};

\draw[-{Latex[length=2mm]}, line width=0.8pt, black!70]
(0.15,-0.55) -- (11.35,-0.55);

\node[font=\scriptsize\bfseries, text=black, anchor=north west] at (0.0,-0.62)
{Lower search cost};

\node[font=\scriptsize\bfseries, text=black, anchor=north east] at (11.5,-0.62)
{Higher QoR / specialization};

\node[font=\scriptsize, text=black] at (5.75,-0.25)
{increasing refinement and optimization intensity};

\end{tikzpicture}
}
\caption{Categorization of representative LLM-based heuristic-generation frameworks into search-efficient, balanced, and high-QoR regimes. Each method is annotated by its core mechanism, showing how different approaches trade off search cost, generality, and optimization quality.}
\label{fig:framework_categories}
\end{figure}
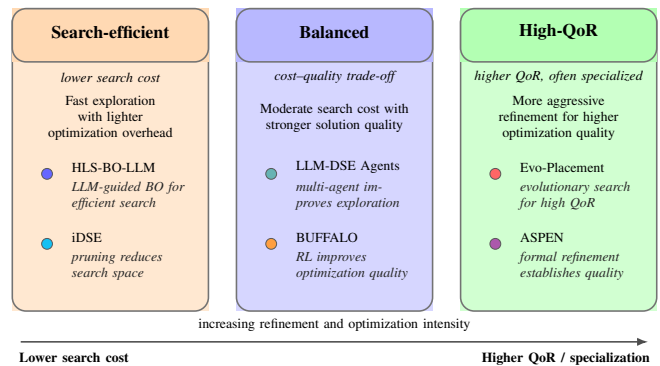

\begin{table*}[t]
\caption{Operational and Qualitative Comparison of Six LLM-Based Heuristic-Generation Frameworks}
\label{tab:comparison}
\centering
\scriptsize
\renewcommand{\arraystretch}{1.5}
\setlength{\tabcolsep}{3pt}
\begin{tabularx}{\textwidth}{>{\RaggedRight}p{1.2cm}>{\RaggedRight}p{1.1cm}>{\RaggedRight}p{1.7cm}>{\RaggedRight}p{1.5cm}>{\RaggedRight}p{1.1cm}>{\RaggedRight}p{1.1cm}>{\RaggedRight}X>{\RaggedRight}X}
\toprule
\textbf{Framework} & \textbf{Design Stage} & \textbf{Core Method} & \textbf{Heuristic Target} & \textbf{Search Cost} & \textbf{Generality} & \textbf{Strengths} & \textbf{Limitations} \\
\midrule
HLS-BO-LLM [1] & HLS DSE & LLM + Bayesian opt. & Directive configs (pragmas) & Low--Medium & Tool-specific & Strong QoR vs. random/grid; interpretable directives & Dependent on surrogate quality and HLS API \\
\midrule
iDSE [2] & HLS DSE & LLM-guided pruning + seeding & Seeds + filters for directives & Low & Moderate & Large reduction in explored designs; convergent/divergent reasoning & Requires tuned prompting and HLS feedback loops \\
\midrule
LLM-DSE Agents [3] & HLS / accelerator params & Multi-agent LLM search & Parameter tree search policy & Medium & Cross-tool potential & Flexible agent roles; good under budgeted evaluations & Coordination overhead; complex to debug \\
\midrule
Evo-Placement [4] & Global placement & LLM-generated alg. variants & Initialization + update rules & Medium--High & Node-specific & Non-trivial HPWL improvements; discovers novel optimizers & Training/evolution cost; integration with legacy placers \\
\midrule
ASPEN [5] & RTL datapath & LLM + e-graphs + formal checks & Rewrite rules and extraction & Medium & Technology-aware & Combines correctness guarantees with PPA gains & E-graph infra and solver overhead; corpus-dependent \\
\midrule
BUFFALO [6] & Buffer-tree / CTS & LLM + GRPO RL & Tree topology and sizing & Low (inference-time) & Industrial-scale & Large timing improvements; orders-of-magnitude runtime gains & Heavy offline training; task-specialized sequence model \\
\bottomrule
\end{tabularx}
\end{table*}

\section{Related Work and Motivation}

Recent research has explored LLMs as active components in optimization workflows across high-level synthesis (HLS), RTL generation, and physical design. Rather than serving as passive code generators, related work positions LLMs as heuristic generators that propose, refine, or adapt optimization strategies under tool feedback \cite{ref1,ref2,ref3}. Figure~1 categorizes existing LLM-based hardware generation approaches into three regimes, namely search-efficient, balanced, and high-QoR, illustrating the trade-offs between optimization quality, search effort, and reliance on tool-driven feedback. Table~\ref{tab:comparison} summarizes the characteristics of related work.

The first category focuses on HLS-driven design space exploration, where heuristics operate over pragma or parameter spaces. Yao et al.~\cite{ref1} integrate an LLM into a Bayesian optimization loop to guide pragma selection. Li et al.~\cite{ref2} extend this idea in iDSE by using the LLM to prune directive spaces, seed promising configurations, and adapt exploration based on observed QoR trends. Wang et al.~\cite{ref3} propose a multi-agent framework in which router, specialist, and critique agents collaboratively traverse parameter trees, with the LLM steering refinement using synthesis feedback. This work showcases that LLMs can reason quite effectively over structured directive spaces when coupled with numerical evaluation signals.

More recent efforts extend LLM-guided heuristics into RTL and physical design, where models generate structured optimization artifacts rather than simple parameter selections. LLMs synthesize program-level constructs such as placement update rules, rewrite sequences, or structural layouts, which are evaluated using synthesis, equivalence checking, and Power-Performance-Area (PPA) feedback. Results show that LLMs can interact meaningfully with downstream EDA tools when constrained to emit executable, tool-evaluable artifacts.

However, prior frameworks also expose persistent limitations\cite{ref5,ref6,ref11}. Many prior approaches couple functional correctness and QoR optimization within a single search loop, wasting effort on unstable designs. Others use weakly regulated exploration, leading to oscillation, premature convergence, and structural drift. In contrast, HYPERHEURIST introduces phase-decoupled correctness gating, stabilizing search by separating correctness discovery from PPA refinement. It models RTL implementations as explicit search states, leverages tool-in-the-loop feedback, and regulates exploration using Simulated Annealing with specialized generation, mutation, and critique roles.

\section{The HYPERHEURIST Framework}

This paper proposes \textit{HYPERHEURIST}, a two-phase heuristic generation framework for RTL code that prioritizes functional correctness before discovering power, performance, and area (PPA) optimization. The central design principle is that correctness and optimization impose fundamentally different search dynamics and should therefore be handled by distinct but coordinated adaptive strategies. By decoupling these objectives, HYPERHEURIST avoids expending expensive optimization effort on invalid designs while allowing aggressive exploration once correctness is established.

\subsection{Framework Architecture}

Figure~2 illustrates the conceptual architecture of the proposed \textsc{HYPERHEURIST} framework.
The framework executes two adaptive strategies sequentially: a correctness-driven search phase followed by a PPA-oriented optimization phase.
We separate these phases because early PPA optimization on partially correct RTL repeatedly produced misleading results in our experiments.

During execution, candidate RTL designs are evaluated using Synopsys VCS simulation \cite{ref20} and synthesis feedback.
The system records the best candidate based on compilation success, simulation pass/fail status, and synthesized area.
Both phases execute feedback-driven search loops, but differ in objective: Phase~1 prioritizes functional validity, while Phase~2 refines power, performance, and area under correctness-preserving constraints.



\subsubsection{Adaptive Strategy 1: Correctness-Driven Search}

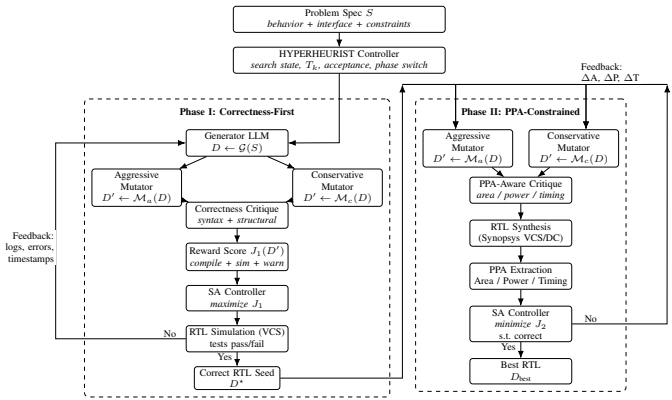
\begin{figure}[t]
\centering
\resizebox{\columnwidth}{!}{%
\begin{tikzpicture}[
  font=\scriptsize,
  node distance=2.6mm and 3.2mm,
  >=Latex,
  box/.style={draw, rounded corners=2pt, align=center, inner sep=2pt,
              minimum width=25mm, minimum height=6mm, fill=white},
  sbox/.style={draw, rounded corners=2pt, align=center, inner sep=1.8pt,
               minimum width=22mm, minimum height=5.6mm, fill=white},
  wide/.style={draw, rounded corners=2pt, align=center, inner sep=2pt,
               minimum width=52mm, minimum height=6mm, fill=white},
  tiny/.style={draw, rounded corners=2pt, align=center, inner sep=1.6pt,
               minimum width=21mm, minimum height=5.2mm, fill=white},
  lab/.style={font=\scriptsize\bfseries, fill=white, inner sep=1.2pt},
  grp/.style={draw, dashed, rounded corners=3pt, inner sep=5pt},
  flow/.style={->, line width=0.45pt},
  fback/.style={->, line width=0.55pt}
]

\node[wide] (spec)
{Problem Spec $S$\\ \emph{behavior + interface + constraints}};

\node[wide, below=3.5mm of spec] (ctrl)
{HYPERHEURIST Controller\\
\emph{search state, $T_k$, acceptance, phase switch}};

\node[lab, below left=8mm and -1mm of ctrl, anchor=north] (p1lab)
{Phase I: Correctness-First};

\node[box, below=3mm of p1lab] (gen)
{Generator LLM\\ $D \leftarrow \mathcal{G}(S)$};

\node[sbox, below left=3mm and 1mm of gen] (amut1)
{Aggressive\\ Mutator\\ $D' \leftarrow \mathcal{M}_a(D)$};

\node[sbox, below right=3mm and 1mm of gen] (cmut1)
{Conservative\\ Mutator\\ $D' \leftarrow \mathcal{M}_c(D)$};

\node[box, below=3.5mm of $(amut1)!0.5!(cmut1)$] (crit1)
{Correctness Critique\\ \emph{syntax + structural}};

\node[box, below=3.5mm of crit1] (score1)
{Reward Score $J_1(D')$\\ \emph{compile + sim + warn}};

\node[box, below=3.5mm of score1] (sa1)
{SA Controller\\ \emph{maximize }$J_1$};

\node[box, below=3.5mm of sa1] (vcs)
{RTL Simulation (VCS)\\ tests pass/fail};

\node[tiny, below=3.5mm of vcs] (seed)
{Correct RTL Seed\\ $D^{\star}$};

\node[grp, fit=(p1lab) (gen) (amut1) (cmut1) (crit1)
                 (score1) (sa1) (vcs) (seed)] (p1grp) {};

\node[lab, right=40mm of p1lab] (p2lab)
{Phase II: PPA-Constrained};

\node[sbox, below=3mm of p2lab, xshift=-13mm] (amut2)
{Aggressive\\ Mutator\\ $D' \leftarrow \mathcal{M}_a(D)$};

\node[sbox, below=3mm of p2lab, xshift=13mm] (cmut2)
{Conservative\\ Mutator\\ $D' \leftarrow \mathcal{M}_c(D)$};

\node[box, below=7mm of $(amut2)!0.5!(cmut2)$] (crit2)
{PPA-Aware Critique\\ \emph{area / power / timing}};

\node[box, below=4mm of crit2] (syn)
{RTL Synthesis\\ (Synopsys VCS/DC)};

\node[box, below=4mm of syn] (ppa)
{PPA Extraction\\ Area / Power / Timing};

\node[box, below=4mm of ppa] (sa2)
{SA Controller\\ \emph{minimize }$J_2$\\ s.t.\ correct};

\node[box, below=4mm of sa2] (best)
{Best RTL\\ $D_{\text{best}}$};

\node[grp, fit=(p2lab) (amut2) (cmut2) (crit2)
                 (syn) (ppa) (sa2) (best)] (p2grp) {};

\coordinate (toplane) at ($(ctrl.north)+(0,14mm)$);
\coordinate (bottomlane) at ($(p1grp.south)+(0,-5mm)$);

\coordinate (ctrl_down) at ($(ctrl.south)+(0,-17mm)$);
\coordinate (gen_up) at ($(gen.north)+(0, 10.5mm)$);

\draw[flow] (spec) -- (ctrl);
\draw[flow] (ctrl.south) -- (ctrl_down) -- (gen.east);

\draw[flow] (gen) -- (amut1);
\draw[flow] (gen) -- (cmut1);
\draw[flow] (amut1) -- (crit1);
\draw[flow] (cmut1) -- (crit1);
\draw[flow] (crit1) -- (score1);
\draw[flow] (score1) -- (sa1);
\draw[flow] (sa1) -- (vcs);
\draw[flow] (vcs) -- node[pos=0.3, left] {Yes} (seed);

\coordinate (fbp1_left) at ($(p1grp.west)+(-7mm,0)$);
\coordinate (fbp1_down) at ($(p1grp.west)+(23mm,-21mm)$);
\coordinate (fbp1_vcs)  at ($(fbp1_left |- vcs)$);
\coordinate (fbp1_gen)  at ($(fbp1_left |- gen)$);

\draw[fback] (vcs.west) -- (fbp1_vcs) -- (fbp1_left) -- (fbp1_gen) -- (gen.west);

\node[left,align=right, fill=white, inner sep=1pt] at (fbp1_left)
{\scriptsize Feedback:\\ logs, errors,\\ timestamps};

\node[left,align=right, fill=white, inner sep=1pt] at (fbp1_down)
{\scriptsize No};

\coordinate (tr_out) at ($(seed.east)+(30mm, 0)$);

\coordinate (tr_trunkx) at ($(p2grp.west)+(-3mm,-12mm)$);

\coordinate (tr_trunk_top) at ($ (15.5mm,-16mm) $);

\coordinate (tr_a) at ($(amut2.north)+(-3mm,0mm)$);
\coordinate (tr_c) at ($(cmut2.north)+(3mm,0mm)$);

\draw[fback] (seed.east) -- (tr_out) -- (tr_trunk_top);

\draw[fback] (tr_trunk_top) -| (tr_a);
\draw[fback] (tr_trunk_top) -| (tr_c);

\draw[flow] (amut2) -- (crit2);
\draw[flow] (cmut2) -- (crit2);
\draw[flow] (crit2) -- (syn);
\draw[flow] (syn) -- (ppa);
\draw[flow] (ppa) -- (sa2);
\draw[flow] (sa2) -- node[pos=0.3, left] {Yes} (best);

\coordinate (p2_right) at ($(p2grp.east)+(10mm,0)$);
\coordinate (fb_out)   at ($(sa2.east)+(3mm,0)$);
\coordinate (fb_right) at ($(p2_right |- sa2)$);
\coordinate (fb_up)    at ($(80.5mm, -16mm)$);

\coordinate (p2grp_left) at ($(p2grp.west)+(45mm, -18mm)$);

\coordinate (fb_to_amut2) at ($(amut2.north)+(-3mm,0mm)$);
\coordinate (fb_to_cmut2) at ($(cmut2.north)+(3mm,0mm)$);
\coordinate (fb_to_tr_trunkx) at ($ (15.5mm,-16mm) $);

\draw[fback] (sa2.east) -- (fb_out) -- (fb_right) -- (fb_up);
\draw[fback] (fb_up) -| (fb_to_amut2);
\draw[fback] (fb_up) -| (fb_to_cmut2);

\node[above,align=left, fill=white, inner sep=1pt] at ($(fb_up)+(-14mm,0mm)$)
{\scriptsize Feedback:\\ $\Delta$A, $\Delta$P, $\Delta$T};

\node[left,align=right, fill=white, inner sep=1pt] at (p2grp_left)
{\scriptsize No};

\end{tikzpicture}%
}
\caption{Two-phase HYPERHEURIST framework. Phase I discovers functionally correct RTL seeds using tool-in-the-loop simulation feedback. Phase II refines the seed via synthesis-driven PPA optimization under a strict correctness constraint.}
\label{fig:hyperheurist_framework}
\end{figure}

This strategy prioritizes discovering functionally valid RTL designs before any optimization is performed. A heuristic generator proposes diverse RTL candidates from the problem specification, emphasizing architectural coverage rather than quality. These candidates are refined through two mutation pathways: a conservative mutator that applies structure-preserving fixes to stabilize existing designs, and an aggressive mutator that introduces larger architectural changes to explore alternative datapaths or control structures.

Each candidate is evaluated by a structural critique that verifies key correctness properties, including reset behavior, assignment discipline, FSM completeness, and pipeline consistency. Compilation, simulation, and structural checks are combined into a scalar correctness reward, which serves as the energy function for a SA controller. This controller probabilistically accepts candidates to balance exploration with convergence toward robust designs. Evaluation feedback is reused to guide subsequent mutations, while the best correctness-preserving candidates are continuously tracked. The output of this phase is a small set of RTL implementations that are functionally verified and structurally stable.

\subsubsection{Adaptive Strategy 2: PPA-Oriented Optimization}

This strategy optimizes power, performance, and area while strictly preserving functional correctness. Mutation and critique retain the same structural roles as in the correctness-driven phase, but mutations are now biased toward synthesis-friendly refinements. Functional correctness is enforced as a hard constraint, and any violating candidate is immediately discarded.

Correct candidates are synthesized to obtain timing, area, and power metrics, which are combined into a composite PPA objective. A SA controller regulates candidate acceptance using this objective, enabling controlled exploration of PPA trade-offs without compromising correctness.

Synthesis feedback is incorporated iteratively, and the framework continuously tracks the best-performing design that remains functionally equivalent to the verified seeds from Phase~1. The output of this phase is a final RTL implementation optimized for PPA under strict correctness constraints.

\subsection{Prompts and Experimental Setup}
\label{sec:prompts_setup}

\subsubsection{Benchmarks and Problem Statements}
\label{sec:benchmarks}
We evaluated \textsc{HYPERHEURIST} using \textbf{8 RTL problem statements} adopted from the \textit{RTLLLM} benchmark set\cite{ref17}. These problem statements are summarized in Table~\ref{tab:benchmarks_table2}. Each benchmark provides a compact natural-language specification and an expected RTL interface, enabling a consistent comparison across baseline prompting and our multi-pipeline hyper-heuristic search.

\begin{table}[t]
\centering
\caption{Eight RTL problem statements adopted from the RTLLLM \cite{ref17} paper}
\label{tab:benchmarks_table2}
\renewcommand{\arraystretch}{1.10}
\setlength{\tabcolsep}{6pt}
\begin{tabular}{p{0.30\linewidth} p{0.62\linewidth}}
\hline
\textbf{Design} & \textbf{Description} \\
\hline
\texttt{serial2parallel\_8} & 1-bit serial input and output data after receiving 6 inputs \\
\texttt{alu4} & Arithmetic logic unit operating on 4-bit inputs \\
\texttt{counter\_0\_12} & Counter module counts from 0 to 12 \\
\texttt{traffic\_light} & Traffic light system with three colors and pedestrian button \\
\texttt{freq\_div} & Frequency divider for 100\,MHz input clock producing lower-frequency outputs \\
\texttt{johnson\_counter} & 4-bit Johnson counter with specific cyclic state sequence \\
\texttt{mux2\_sync} & Multi-bit synchronous multiplexer \\
\texttt{parallel2serial} & Convert 4 input bits into a single serial output bit \\
\hline
\end{tabular}
\end{table}


\subsubsection{Multi-Pipeline Prompting (Four LLM Roles)}
\label{sec:prompting_pipelines}
For each problem statement, \textsc{HYPERHEURIST} initialized a candidate pool using four distinct LLM pipelines (roles). The goal was to induce complementary behaviors (diverse exploration vs.\ safe refinement) while maintaining a consistent output contract (synthesizable RTL).

\begin{itemize}
    \item \textbf{Pipeline A: Generator} produces a clean, standard RTL implementation targeting correctness.
    \item \textbf{Pipeline B: Conservative Mutator} makes minimal, low-risk edits (small rewrites, reset fixes, interface consistency).
    \item \textbf{Pipeline C: Critique} reviews the candidate RTL and emits actionable defect hypotheses and patch suggestions.
    \item \textbf{Pipeline D: Aggressive Mutator} performs larger transformations (state encoding changes, pipelining, refactoring) to escape local minima.
\end{itemize}

\subsubsection{Common System Contract}
All pipelines share a strict system-level contract to ensure compilation and synthesis compatibility.

\vspace{4pt}
\noindent
{\ttfamily\small
\textbf{SYSTEM\_BASE:}\\
You are an expert RTL engineer.\\
Return ONLY synthesizable SystemVerilog code (no markdown, no explanation).\\
Keep module name and ports exactly as specified.
}

\subsubsection{Role-Specific Prompt Templates}
Each pipeline receives the same benchmark specification but is conditioned using a role-specific prompt template. Curly-brace fields denote runtime substitution during execution.

\vspace{6pt}
\noindent
{\ttfamily\small
\textbf{GENERATOR\_PROMPT:}\\
You are a RTL engineer.\\
TASK: Parameterizable Johnson counter (W=8): if \texttt{ce}, shift and set \texttt{q[0]\texttt{<=}\textasciitilde q[W-1]}. Add sync active-high reset to a known state + illegal-state recovery to reset state.\\
CONSTRAINTS: Synthesizable SystemVerilog; \texttt{always\_ff}/\texttt{always\_comb} separated; fully synchronous; all regs reset.\\
OUTPUT: Synthesizable SystemVerilog only.
}

\vspace{6pt}
\noindent
{\ttfamily\small
\textbf{CONSERVATIVE\_MUTATOR\_PROMPT:}\\
REF: \{rtl\}\\
TASK: Fix correctness/synth/lint issues without changing architecture; keep reset/enable behavior consistent; confirm illegal-state recovery.\\
CONSTRAINTS: Keep ports/params and pipeline depth unchanged; synchronous reset.\\
OUTPUT: Synthesizable SystemVerilog only.
}

\vspace{6pt}
\noindent
{\ttfamily\small
\textbf{AGGRESSIVE\_MUTATOR\_PROMPT:}\\
REF: \{rtl\}\\
TASK: Explore alternate micro-architecture (e.g., optional staging/enable pipelining) while preserving spec and recovery behavior.\\
CONSTRAINTS: Keep ports/params unchanged; synchronous reset; synthesizable SystemVerilog.\\
OUTPUT: Synthesizable SystemVerilog only.
}

\vspace{6pt}
\noindent
{\ttfamily\small
\textbf{CRITIQUE\_PROMPT:}\\
SPEC: Johnson counter (W=8) with \texttt{ce}, inverted-MSB feedback, sync active-high reset, and illegal-state recovery.\\
RTL: \{rtl\}\\
OUTPUT: JSON only: \{syntax, reset, logic, hazard\} in \{0.0,0.5,1.0\}.
}

\subsubsection{Tool-in-the-Loop Verification and Feedback}
\label{sec:tools_feedback}
We compiled and simulated each candidate using \texttt{Synopsys VCS W-2024.09-SP1\_Full64} \cite{ref20}. If a candidate failed at any stage, the tool-generated error outputs were fed back into the next LLM call to drive targeted repair. The feedback packet included: (i) compile errors, (ii) simulation failures, (iii) warnings, and (iv) timestamps/error identifiers when available.

{\em Compile/Simulation Loop.}
Given candidate RTL $x$, we ran:
\begin{enumerate}
    \item \textbf{Compile check:} validate syntax and elaboration.
    \item \textbf{Simulation check:} execute the benchmark testbench and record pass/fail.
    \item \textbf{Log feedback:} on failure, extract a compact log slice (first error + context) and return it to the selected pipeline for repair.
\end{enumerate}



\begin{table}[t]
\centering
\caption{Syntax and Functional Correctness Comparison Across LLM-Based RTL Generation Frameworks}
\label{tab:correctness_detailed}
\renewcommand{\arraystretch}{1.1}
\setlength{\tabcolsep}{3pt}
\resizebox{\columnwidth}{!}{%
\begin{tabular}{|l|cc|cc|cc|}
\hline
\rowcolor{blue!25}
\cellcolor{blue!25} & \multicolumn{4}{c|}{\textbf{RTL--LLM [13]}} & \multicolumn{2}{c|}{\textbf{HYPERHEURIST}} \\
\cline{2-7}
\rowcolor{blue!15}
\multicolumn{1}{|c|}{\cellcolor{blue!25}\multirow{-2}{*}[1.8pt]{\textbf{Design}}} & \multicolumn{2}{c|}{\textbf{GPT-3.5 + SP}} & \multicolumn{2}{c|}{\textbf{GPT-4}} & \multicolumn{2}{c|}{\textbf{GPT-4.0}} \\
\cline{2-7}
\rowcolor{blue!10}
\cellcolor{blue!25} & \textbf{Syn} & \textbf{Func} & \textbf{Syn} & \textbf{Func} & \textbf{Syn} & \textbf{Func} \\
\hline
serial2parallel\_8 & \cmark & \cmark & \cmark & \cmark & \cmark & \cmark \\
alu4               & \xmark & \xmark & \cmark & \xmark & \cmark & \cmark \\
counter\_0\_12     & \cmark & \cmark & \cmark & \cmark & \cmark & \cmark \\
traffic\_light     & \cmark & \xmark & \cmark & \cmark & \xmark & \xmark \\
freq\_div          & \cmark & \xmark & \cmark & \xmark & \cmark & \cmark \\
johnson\_counter   & \cmark & \cmark & \cmark & \cmark & \cmark & \cmark \\
mux2\_sync         & \cmark & \xmark & \cmark & \cmark & \cmark & \cmark \\
parallel2serial    & \cmark & \xmark & \cmark & \xmark & \cmark & \xmark \\
\hline
\end{tabular}%
}
\end{table}

HYPERHEURIST uses the same Simulated Annealing (SA) control logic in both phases. The only difference is the \emph{objective function} used to score a design and the \emph{constraints} enforced by the evaluator:
(i) Phase 1 optimizes correctness, and
(ii) Phase 2 optimizes PPA while preserving correctness.
Algorithm 1 presents the unified SA routine.

\subsection{Unified Simulated Annealing Procedure for Phase 1 and Phase 2}
\label{sec:unified_sa_algo}

\begin{algorithm}[t]
\caption{Unified Simulated Annealing for HYPERHEURIST (Phase 1 \& Phase 2)}
\label{alg:unified_sa_hyperheurist}
\small
\begin{algorithmic}[1]
\Require Benchmark specification $S$; initial design $D_0$; max iterations $K$; initial temperature $T_0$; cooling factor $\alpha\in(0,1)$; minimum temperature $T_{\min}$; mode schedule $\textsc{Mode}(k)\in\{\textsc{P1},\textsc{P2}\}$.
\Ensure Best design found $D_{\text{best}}$ and score $J_{\text{best}}$.

\State $D \gets D_0$
\State $(J, \textsc{Logs}) \gets \textsc{Evaluate}(D,S,\textsc{Mode}(0))$
\State $D_{\text{best}} \gets D,\quad J_{\text{best}} \gets J$
\State $T \gets T_0$

\For{$k=1$ to $K$}
  \If{$T < T_{\min}$}
    \State \textbf{break}
  \EndIf

  \State $m \gets \textsc{Mode}(k)$

  \Comment{LLM-guided neighbor generation may use feedback logs}
  \State $D' \gets \textsc{MutateLLM}(D,S,m,\textsc{Logs})$

  \State $(J', \textsc{Logs}') \gets \textsc{Evaluate}(D',S,m)$

  \Comment{Unified SA acceptance on scalar objective $J$ (higher is better)}
  \State $\Delta \gets J' - J$
  \If{$\Delta \ge 0$}
      \State $D \gets D',\; J \gets J',\; \textsc{Logs} \gets \textsc{Logs}'$
  \Else
      \State accept with probability $p=\exp(\Delta/T)$
      \If{accepted}
          \State $D \gets D',\; J \gets J',\; \textsc{Logs} \gets \textsc{Logs}'$
      \EndIf
  \EndIf

  \If{$J > J_{\text{best}}$}
    \State $D_{\text{best}} \gets D,\quad J_{\text{best}} \gets J$
  \EndIf

  \State $T \gets \alpha\cdot T$
\EndFor

\State \Return $D_{\text{best}}, J_{\text{best}}$
\end{algorithmic}
\end{algorithm}

\subsubsection{How the Algorithm Works in Both Phases}

Both phases share the same algorithmic structure; they differ only in the evaluation objective and acceptance rule.

\begin{itemize}
    \item \textbf{Phase 1 (Correctness SA):} \textsc{Evaluate} maximizes a correctness score. Candidates with $\Delta \ge 0$ are always accepted, while worse candidates are accepted with probability $\exp(\Delta/T)$ to escape partially correct RTL plateaus.

    \item \textbf{Phase 2 (PPA SA):} \textsc{Evaluate} minimizes a PPA cost. Candidates with $\Delta \le 0$ are always accepted, while degradations are accepted with probability $\exp(-\Delta/T)$ to explore beyond local optima.

    \item \textbf{Mutation and constraints:} \textsc{MutateLLM} is guided by tool feedback, compilation and simulation logs in Phase~1 \cite{ref12,ref13}, and synthesis, timing, and power reports in Phase~2, while Phase~2 mutations strictly preserve functional correctness.

    \item \textbf{Temperature schedule:} High temperature promotes exploration, with gradual cooling enforcing convergence.
\end{itemize}


\section{Experimental Controls and Reproducibility}
\label{sec:controls_repro}

All HYPERHEURIST correctness and PPA results (Tables IV–V) were obtained under controlled conditions, while baseline correctness results in Table III are reproduced from prior work \cite{ref17}. Each framework was evaluated using the same toolchain, constraints, and verification flow so that observed differences reflect only the generation strategy.

\paragraph{Controlled tool-in-the-loop evaluation}
For each benchmark, RTL candidates were generated automatically and evaluated using a fixed two-stage pipeline:
\begin{enumerate}
\item Front-end correctness validation via compilation and simulation using \textit{Synopsys VCS} \cite{ref20}, enforcing consistent testbench execution and deterministic pass/fail outcomes.
\item Back-end PPA evaluation via logic synthesis using \textit{Synopsys Design Compiler} (DC) \cite{ref20} with a fixed standard-cell library and identical timing constraints across all benchmarks.
.
\end{enumerate}

\begin{lstlisting}[style=svsuccess, caption={Correct and stable RTL for \texttt{freq\_div}}]
module freq_div #(parameter DIV = 100)(
  input  logic clk,
  input  logic rst_n,
  output logic clk_out
);
  logic [$clog2(DIV)-1:0] cnt;

  always_ff @(posedge clk) begin
    if (!rst_n) begin
      cnt <= '0;              // Proper reset
      clk_out <= 1'b0;        // Deterministic initialization
    end else if (cnt == DIV-1) begin
      cnt <= '0;
      clk_out <= ~clk_out;    // Controlled toggle
    end else begin
      cnt <= cnt + 1'b1;
    end
  end
endmodule
\end{lstlisting}

This design synthesizes efficiently due to minimal control logic and predictable switching activity and leads to reduced area and power in Table~IV.
No manual edits, constraint tuning, or post-processing were applied.

Let $N$ denote the total number of generated candidates. For candidate $i$, define:
\begin{equation}
\mathbb{I}_{\text{vcs}}^{(i)} =
\begin{cases}
1 & \text{if VCS compilation and simulation succeed} \\
0 & \text{otherwise}
\end{cases}
\end{equation}

Only candidates with $\mathbb{I}_{\text{vcs}}^{(i)} = 1$ were forwarded to synthesis, ensuring that all reported PPA results correspond to functionally valid RTL.

\paragraph{PPA objective formulation.}
For each validated candidate, DC reports:
\[
A^{(i)}, \quad P^{(i)}, \quad T^{(i)}.
\]
The final design is selected by minimizing:
\begin{equation}
\mathcal{J}^{(i)} = \alpha \cdot \widehat{A}^{(i)} + \beta \cdot \widehat{P}^{(i)} + \gamma \cdot \widehat{T}^{(i)},
\label{eq:ppa_cost}
\end{equation}
where $\widehat{(\cdot)}$ denotes per-benchmark normalization and $\alpha,\beta,\gamma$ are fixed weights. The objective is evaluated over the feasible set:
\[
\mathcal{F} = \{ i \mid \mathbb{I}_{\text{vcs}}^{(i)} = 1 \},
\]
and the reported design is $i^\star = \arg\min_{i \in \mathcal{F}} \mathcal{J}^{(i)}$.

\paragraph{Representative successful case.}
The \texttt{freq\_div} benchmark illustrates a successful outcome. All generated candidates pass VCS validation, and the simple synchronous counter structure supports stable heuristic refinement.

\paragraph{Failure case study: \texttt{traffic\_light}.}
The \texttt{traffic\_light} benchmark represents a controlled failure case. All generated candidates fail during VCS simulation and are therefore excluded from PPA evaluation.

\begin{lstlisting}[style=svfailure, caption={Semantically incorrect RTL fragment for \texttt{traffic\_light}}]
always_ff @(posedge clk) begin
  if (rst) begin
    state <= RED;
  end else begin
    case (state)
      RED:    if (timer == 0) state <= GREEN;
      GREEN:  if (timer == 0) state <= YELLOW;
      YELLOW: if (timer == 0) state <= RED;
      // Missing default case
    endcase
    // Missing timer reset / update
  end
end
\end{lstlisting}

The failure is caused by incomplete timer semantics and missing default assignments, leading to simulation-time assertion failures. Because $\mathbb{I}_{\text{vcs}}^{(i)} = 0$ for all candidates, the design is never forwarded to DC, explaining the absence of PPA results for \texttt{traffic\_light} in Table~IV.

\paragraph{Reproducibility guarantees.}
All experiments use
fixed random seeds, deterministic prompt templates and version-pinned EDA tools. Scripts, RTL artifacts, synthesis logs, and VCS reports are archived, ensuring that reported correctness and PPA trends are reproducible across independent
runs. All evaluation pipelines are fully automated to avoid manual intervention. Tool invocations, command-line options, and environment variables are logged to enable exact re-execution of the full flow.


\lstdefinestyle{compilerlog}{
  basicstyle=\ttfamily\footnotesize,
  columns=fullflexible,
  keepspaces=true,
  breaklines=true,
  frame=single,
  rulecolor=\color{black},
  showstringspaces=false,
  aboveskip=6pt,
  belowskip=6pt
}

\subsection{End-to-End HYPERHEURIST Optimization Trace}
\label{sec:p1_p2_trace}

Listing~3 presents an end-to-end execution of the \textbf{HYPERHEURIST} framework on the \texttt{johnson\_counter} benchmark, covering both Phase~1 (correctness-driven search using VCS) and Phase~2 (PPA-driven optimization using Design Compiler). Only RTL candidates that pass functional verification are forwarded for synthesis evaluation. Iterative LLM-guided mutations progressively refine the design while maintaining correctness, and multiple valid candidates are compared based on area, timing, and power trade-offs. A unified simulated-annealing schedule enables controlled design-space exploration, while detailed iteration logs ensure traceability and reproducibility. Overall, the listing demonstrates the feasibility of integrating verification and synthesis feedback within a unified closed-loop RTL optimization flow.

\vspace{-4pt}
\begin{lstlisting}[
style=hhlog,
caption={Representative Phase--1 (Correctness SA) and Phase--2 (PPA SA) optimization trace for \texttt{johnson\_counter}.},
label={lst:p1p2_compact}
]
=== PHASE 1: Correctness SA ===
[P1] iter=0   T=1.20  score=0.56  compile=1 sim=1   ACCEPT
[P1] iter=4   T=0.38  score=0.98  compile=1 sim=1   ACCEPT
[P1] iter=6   T=0.21  score=0.78  compile=0 sim=0   REJECT
[P1] iter=5   T=0.29  score=0.99  compile=1 sim=1   SELECTED
[P1] best_score=0.99  -> out_phase1_best.sv

=== PHASE 2: PPA SA (DC) ===
[P2] iter=6   T=0.26  area=64.3  power=92.7  wns=0.182  ACCEPT
[P2] iter=7   T=0.21  area=61.8  power=85.4  wns=0.196  ACCEPT
[P2] iter=10  T=0.11  area=66.2  power=95.8  wns=0.175  REJECT
[P2] iter=8   T=0.17  area=59.9  power=80.9  wns=0.200  SELECTED
[P2] best_score=2.90e-01 -> out_phase2_best.sv
\end{lstlisting}

\vspace{-6pt}
\begin{center}
\footnotesize
\textcolor{LogGreen}{\textbf{ACCEPT}} indicates an accepted SA move;\quad \textcolor{LogRed}{\textbf{REJECT}} indicates rejection under the Metropolis criterion;\quad
\textcolor{LogPurple}{\textbf{SELECTED}} marks the final PPA point reported in Table~IV.
\end{center}
\vspace{-4pt}

\paragraph{Discussion.}
Phase~1 converges toward a fully correct RTL candidate under strict
compile and simulation gates, while Phase~2 performs localized,
temperature-controlled exploration within the feasible design space.
The selected solution achieves the lowest normalized PPA objective
while maintaining positive timing slack, directly corresponding to
the values reported for \texttt{johnson\_counter} in Table~IV.

\subsection{PPA Measurement Using Synopsys Design Compiler}

Phase--2 PPA evaluation is performed using Synopsys Design Compiler (DC) V-2023.12-SP5 \cite{ref20} under a fixed technology and constraint setup. All RTL candidates are synthesized using the same 90\,nm NAND-gate standard-cell library (\texttt{.db}), ensuring that reported PPA differences reflect only RTL and architectural changes introduced during LLM-driven mutation and simulated annealing (SA).

Each candidate is synthesized using a deterministic DC script that reads and elaborates the RTL, links against the 90\,nm library, applies uniform timing constraints, performs technology mapping, and reports timing, area, and power. Identical clock definitions, I/O delays, clock uncertainty, and load models are used across all candidates.

\textbf{Metrics:}
Timing is reported as the worst negative slack (WNS) in nanoseconds
from \texttt{report\_timing}. Area is obtained from \texttt{report\_area},
and total power is computed from \texttt{report\_power} as
\begin{equation}
P_{\mathrm{total}} = P_{\mathrm{leak}} + P_{\mathrm{internal}} + P_{\mathrm{switch}} .
\end{equation}
Identical activity assumptions are used across all designs.

\textbf{PPA objective:}
To guide Phase--2 SA selection, DC-reported metrics are combined into a normalized objective:
\begin{equation}
  J_{\mathrm{PPA}} \;=\; w_A \,\widehat{A} \;+\; w_P \,\widehat{P} \;+\; w_S \,\widehat{S},
\end{equation}
where $\widehat{A}$ and $\widehat{P}$ are normalized area and power, and $\widehat{S}$ penalizes timing violations (e.g., negative slack). The selected Phase--2 design minimizes $J_{\mathrm{PPA}}$ under the Metropolis acceptance rule while satisfying all Phase--1 correctness checks. The resulting PPA values are reported in Table~IV.

\subsection{Run Protocol}

For each problem $P_i$ and each method (baseline vs. \textsc{HyperHeurist}), we run multiple independent trials:

\begin{itemize}
\item \textbf{Baseline}: GPT-4.0 is queried five times with the generator prompt only; each run returns a single candidate RTL.
\item \textbf{\textsc{HyperHeurist}}: For each $P_i$, five runs are performed. In each run, the bandit is initialized with a uniform prior over pipelines and allowed a fixed budget of generator/mutator--critique cycles. After each candidate, the critique scores are combined into $R$, and the bandit updates its parameters.
\end{itemize}

Within a run, the final design for $P_i$ is the candidate with highest reward $R$ that also satisfies $S_{\text{syntax}} = 1$. Across runs, we compute the empirical success rates reported in Table IV (syntax correctness, structural correctness, and qualitative heuristic depth).
\section{Results and Discussion}
\label{sec:results}

This section evaluated the behavior of \textsc{HYPERHEURIST} on representative RTL benchmarks, focusing on correctness convergence, PPA refinement, and observed failure cases. The results showed that separating correctness discovery from PPA optimization leads to stable and repeatable improvements over baseline LLM-based RTL generation.

\subsection{Correctness Convergence}

In all benchmarks that complete Phase~1, \textsc{HYPERHEURIST} converged to functionally correct RTL under strict compilation and simulation checks using Synopsys VCS. As summarized in Table~V, the framework consistently improved structural and logical correctness compared to baseline LLM outputs. The Phase~1 SA trace (Listing~1) illustrates this process. Designs that failed compilation or simulation were immediately rejected, preventing invalid candidates from influencing later stages. For the \texttt{johnson\_counter} benchmark, the correctness score steadily increased and stabilized near $0.99$, indicating convergence to a valid and stable RTL implementation. This behavior differs from single-pass LLM generation, where correctness is largely dependent on prompt quality and lacks recovery from tool feedback.

\begin{table*}[t]
\centering
\caption{PPA Comparison of Gate-Level Netlists Synthesized with Synopsys Design Compiler}
\label{tab:ppa_comparison}

\renewcommand{\arraystretch}{1.2}
\small

\resizebox{\textwidth}{!}{%

\rowcolors{3}{gray!10}{white}
\begin{tabular}{|l|c|c|c||c|c|c||c|c|c|}
\hline

\rowcolor{blue!25}
\textbf{Design}
& \multicolumn{3}{c||}{\textbf{RTL-LLM \cite{ref17} (ChatGPT-4.0)}}
& \multicolumn{3}{c||}{\textbf{RTL-LLM \cite{ref17} (GPT-3.5 + SP)}}
& \multicolumn{3}{c|}{\textbf{HYPERHEURIST (GPT-4.0)}} \\
\hline

\rowcolor{blue!15}
& \textbf{Area ($\mu$m$^2$)}
& \textbf{Power ($\mu$W)}
& \textbf{Timing (ns)}
& \textbf{Area ($\mu$m$^2$)}
& \textbf{Power ($\mu$W)}
& \textbf{Timing (ns)}
& \textbf{Area ($\mu$m$^2$)}
& \textbf{Power ($\mu$W)}
& \textbf{Timing (ns)} \\
\hline

serial2parallel\_8
& 100 & 9800 & \textcolor{red!70!black}{-0.28}
& 155 & 14000 & \textcolor{red!70!black}{-0.33}
& \textbf{125.9} & \textbf{140.9} & \textcolor{green!60!black}{\textbf{+0.39}} \\
\hline

alu4
& 3300 & 1400 & \textcolor{red!70!black}{-0.71}
& -- & -- & --
& \textbf{201.5} & \textbf{156.8} & \textcolor{green!60!black}{\textbf{+0.28}} \\
\hline

counter\_0\_12
& 46 & 4400 & \textcolor{red!70!black}{-0.26}
& 76 & 8400 & \textcolor{red!70!black}{-0.26}
& \textbf{43.6} & \textbf{45.6} & \textcolor{green!60!black}{\textbf{+0.35}} \\
\hline

traffic\_light
& 138 & 11000 & \textcolor{red!70!black}{-0.38}
& -- & -- & --
& -- & -- & -- \\
\hline

freq\_div
& 118 & 16000 & \textcolor{red!70!black}{-0.32}
& 667 & 53000 & \textcolor{red!70!black}{-0.41}
& 322.0 & 185.6 & \textcolor{green!60!black}{\textbf{+0.04}} \\
\hline

johnson\_counter
& 42 & 4700 & \textcolor{red!70!black}{-0.26}
& 195 & 21000 & \textcolor{red!70!black}{-0.22}
& 59.9 & \textbf{80.9} & \textcolor{green!60!black}{\textbf{+0.20}} \\
\hline

mux2\_sync
& 90 & 9.5 & \textcolor{red!70!black}{-0.08}
& 144 & 14 & \textcolor{red!70!black}{-0.08}
& \textbf{7.05} & \textbf{6.45} & \textcolor{green!60!black}{\textbf{+0.38}} \\
\hline

parallel2serial
& 20 & 3800 & \textcolor{red!70!black}{-0.19}
& 1.06 & 0 & 0
& -- & -- & -- \\
\hline

\end{tabular}
}
\end{table*}

\subsection{PPA Optimization}

After correctness was established, Phase~2 performed localized exploration of the RTL space using Design Compiler feedback. Final PPA results were reported in Table~IV.

For the \texttt{johnson\_counter} benchmark, \textsc{HYPERHEURIST} achieved an area of $59.9~\mu m^2$, power of $80.9~\mu W$, and a positive timing slack of $+0.20~ns$. The Phase~2 trace showed that this solution was reached through gradual SA refinement rather than isolated selection. Intermediate candidates with higher area or power were rejected as the temperature decreases, indicating convergence to a stable optimum.

The Phase~2 objective was defined as:
\begin{equation}
\mathcal{J} = \alpha \widehat{A} + \beta \widehat{P} + \gamma \widehat{T},
\end{equation}
where $\widehat{A}$, $\widehat{P}$, and $\widehat{T}$ are normalized area, power, and timing metrics. By optimizing $\mathcal{J}$ only over VCS-validated designs, the framework avoids trading correctness for QoR.

\begin{table*}[t]
\centering
\caption{Correctness improvement of \textsc{HYPERHEURIST} over baseline RTL generation}
\label{tab:correctness_metrics}

\renewcommand{\arraystretch}{1.2}
\setlength{\tabcolsep}{5pt}
\small

\rowcolors{3}{gray!10}{white}

\begin{tabular}{|l|c|c|c||c|c|c||c|c|c|}
\hline

\rowcolor{blue!25}
\textbf{Design} &
\multicolumn{3}{c||}{\textbf{Baseline}} &
\multicolumn{3}{c||}{\textbf{HYPERHEURIST}} &
\textbf{Structural} &
\textbf{Relative} &
\textbf{Depth} \\

\rowcolor{blue!15}
&
\textbf{Syntax} &
\textbf{Structural} &
\textbf{Logic} &
\textbf{Syntax} &
\textbf{Structural} &
\textbf{Logic} &
\textbf{$\Delta$} &
\textbf{Gain} &
\\
\hline

serial2parallel\_8 & 85 & 55 & 55 & 86 & 90 & 90 & \textcolor{green!60!black}{+35} & \textcolor{green!60!black}{+63.6} & Medium \\
\hline
alu4 & 90 & 60 & 60 & 89 & 88 & 88 & \textcolor{green!60!black}{+28} & \textcolor{green!60!black}{+46.7} & Medium \\
\hline
counter\_0\_12 & 95 & 75 & 75 & 96 & 95 & 95 & \textcolor{green!60!black}{+20} & \textcolor{green!60!black}{+26.7} & Low \\
\hline
traffic\_light & 80 & 35 & 35 & 78 & 0 & 0 & \textcolor{red!70!black}{$-35$} & \textcolor{red!70!black}{$-100.0$} & Medium \\
\hline
freq\_div & 88 & 50 & 50 & 89 & 85 & 85 & \textcolor{green!60!black}{+35} & \textcolor{green!60!black}{+70.0} & High \\
\hline
johnson\_counter & 92 & 70 & 70 & 93 & 93 & 93 & \textcolor{green!60!black}{+23} & \textcolor{green!60!black}{+32.9} & Low--Medium \\
\hline
mux2\_sync & 98 & 90 & 90 & 98 & 98 & 98 & \textcolor{green!60!black}{+8} & \textcolor{green!60!black}{+8.9} & Low \\
\hline
parallel2serial & 82 & 40 & 40 & 81 & 0 & 0 & \textcolor{red!70!black}{$-40$} & \textcolor{red!70!black}{$-100.0$} & Medium \\
\hline

\end{tabular}
\end{table*}

\subsection{Comparison with Baselines}

As shown in Table~IV, baseline RTL-LLM \cite{ref17} approaches (ChatGPT-4.0 and GPT-3.5 with self-planning) often produce designs with higher area and power or unstable timing behavior. In contrast, \textsc{HYPERHEURIST} consistently yields compact implementations with lower power and positive slack, particularly for structured designs such as \texttt{johnson\_counter} and \texttt{mux2\_sync}. These gains arise from the framework structure rather than model scale. Multiple generation and mutation pipelines, combined with tool-driven feedback and SA-based acceptance, enable systematic refinement that single-shot approaches lack.

\subsection{Failure Cases}

Some benchmarks did not succeed. The \texttt{traffic\_light} design failed in Phase~1 due to semantic errors in state transitions and reset behavior. These issues were detected during VCS simulation and filtered before synthesis; hence, no PPA results are reported in Table~IV. This behavior is intentional, as invalid RTL is rejected early to avoid misleading PPA evaluation. Baseline RTL--LLM PPA values are reproduced from the RTLLLM benchmark study~\cite{ref17} without modification; near-zero power or area values reflect synthesis optimizations reported in that work.

\subsection{Correctness Evaluation and Metrics}

Table~\ref{tab:correctness_metrics} summarizes correctness results for the baseline LLM pipeline and \textsc{HYPERHEURIST} across all RTL benchmarks. Correctness was evaluated along three dimensions: \emph{syntax correctness}, which measures whether generated RTL parses and compiles successfully; \emph{structural correctness}, which captures compliance with architectural and interface constraints such as reset behavior, state encoding, and module connectivity; and \emph{logic correctness}, which reflects functional validity under simulation. Logic correctness was evaluated only for structurally valid designs and therefore coincides with structural correctness in this study.

Let $N$ denote the total number of generated RTL candidates for a benchmark. Syntax correctness is defined as
\begin{equation}
S_{\text{syntax}} = \frac{N_{\text{syntax-pass}}}{N} \times 100,
\end{equation}
while structural correctness is given by
\begin{equation}
S_{\text{struct}} = \frac{N_{\text{struct-pass}}}{N} \times 100.
\end{equation}
Since logic correctness is assessed only for structurally valid designs,
\begin{equation}
S_{\text{logic}} = S_{\text{struct}}.
\end{equation}
To quantify improvement over the baseline, we report the absolute structural gain
\begin{equation}
\Delta S_{\text{struct}} = S_{\text{struct}}^{\text{Hyper}} - S_{\text{struct}}^{\text{Base}},
\end{equation}
as well as the relative gain
\begin{equation}
G_{\text{rel}} =
\frac{S_{\text{struct}}^{\text{Hyper}} - S_{\text{struct}}^{\text{Base}}}
{S_{\text{struct}}^{\text{Base}}}
\times 100.
\end{equation}

To characterize the complexity of corrective transformations applied during generation, we define a heuristic depth score
\begin{equation}
R = \alpha_1 S_{\text{syntax}} 
  + \alpha_2 S_{\text{reset}} 
  + \alpha_3 S_{\text{pipeline}} 
  + \alpha_4 S_{\text{logic}} 
  + \alpha_5 S_{\text{hazard}},
\end{equation}
where $\alpha_i$ are tunable weights and each term represents the fraction of candidates satisfying the corresponding constraint. Benchmarks are categorized as \emph{Low}, \emph{Medium}, or \emph{High} heuristic depth based on $R$.

Across the evaluated benchmarks, \textsc{HYPERHEURIST} improved structural correctness on six of eight designs, with absolute gains ranging from $+8\%$ to $+35\%$ and relative improvements reaching up to $70\%$. Syntax correctness remained comparable to the baseline, indicating that observed gains arise from structural refinement rather than syntactic repair. Benchmarks classified as \emph{High} heuristic depth exhibited the largest improvements, while \emph{Low} depth designs showed smaller but consistent gains. Two benchmarks (\texttt{traffic\_light} and \texttt{parallel2serial}) exhibited negative deltas. These designs are dominated by tightly coupled FSM logic with strict multi-cycle temporal dependencies. Failures occurred during VCS simulation due to assertion violations, and the limited diagnostic information available from simulation logs restricts effective corrective feedback. While the feedback-driven regeneration loop converged for arithmetic and control-light designs, these control-intensive cases highlight current limits of heuristic intervention under strict temporal constraints.

\section{Conclusion}

This paper presented \textsc{HYPERHEURIST}, a simulated annealing–based control framework that integrates large language models into RTL design as heuristic generators rather than final code producers. By embedding LLM-generated candidates within a tool-driven optimization loop, the framework enables iterative refinement under real compilation, simulation, and synthesis feedback. A key design principle is prioritizing functional correctness before performance optimization, avoiding wasted effort on invalid designs and yielding a more stable and reproducible search trajectory. Experimental results demonstrate consistent improvements in correctness convergence, along with PPA gains of up to 70\%, highlighting the effectiveness of phase-decoupled exploration. The framework also enhances design traceability through structured optimization logs and supports controlled exploration of the RTL design space with fewer structural regressions. Future work includes extending the approach to larger control-intensive systems (e.g., pipeline controllers and complex FSMs) and incorporating feedback from downstream physical design stages such as placement and routing, further improving scalability and practical integration into industrial RTL flows.

\end{document}